\documentclass[aps,prl,amsmath,amssymb,draft]{revtex4}
\pdfoutput=1
\usepackage{amsthm}
\usepackage{graphicx}
\usepackage{color}
\newcommand{\beq}{\begin{eqnarray}}
\newcommand{\eeq}{\end{eqnarray}}

\newcommand{\bmp}{\noindent\begin{minipage}{16cm}}
\newcommand{\emp}{\end{minipage}\vskip 7mm} 

\usepackage{dcolumn}
\usepackage{bm}
\usepackage{bbm}
\usepackage{subfigure}
\usepackage{pxfonts}

\theoremstyle{definition}

\theoremstyle{plain}

\usepackage{epsfig}

\usepackage[ margin=5pt, font=normalsize,labelfont=bf,justification=raggedright]{caption}
\usepackage{youngtab}
\usepackage{slashed}

\definecolor{rossoCP3}{cmyk}{0,.88,.77,.40}

\def\lsim{\mathrel{\rlap{\lower4pt\hbox{\hskip1pt$\sim$}}
    \raise1pt\hbox{$<$}}}                
\def\gsim{\mathrel{\rlap{\lower4pt\hbox{\hskip1pt$\sim$}}
    \raise1pt\hbox{$>$}}}                

\newcommand{\drawsquare}[2]{\hbox{%
\rule{#2pt}{#1pt}\hskip-#2pt
\rule{#1pt}{#2pt}\hskip-#1pt
\rule[#1pt]{#1pt}{#2pt}}\rule[#1pt]{#2pt}{#2pt}\hskip-#2pt
\rule{#2pt}{#1pt}}

\newcommand{\Yfund}{\raisebox{-.5pt}{\drawsquare{6.5}{0.4}}}

\begin{document}
\title{\Large  \color{rossoCP3} The Standard Model is Natural \\ as \\ Magnetic Gauge Theory}
\author{Francesco Sannino$^{\color{rossoCP3}{\varheartsuit}}$}\email{sannino@cp3.sdu.dk} 
\affiliation{
$^{\color{rossoCP3} {\varheartsuit}}${ CP}$^{ \bf 3}${-Origins}, 
University of Southern Denmark, Campusvej 55, DK-5230 Odense M, Denmark.
}
\begin{abstract}
We suggest  that the Standard Model can be viewed as the magnetic dual of a gauge theory featuring only fermionic matter content. We show this by first introducing a Pati-Salam like extension of the Standard Model and then relating it to a possible dual electric theory featuring only fermionic matter. The absence of scalars in the electric theory indicates that the associated magnetic theory is free from quadratic divergences. Our novel solution to the Standard Model hierarchy problem leads also to a new insight on the mystery of the observed number of fundamental fermion generations by naturally explaining why it has to be at least three.
\\[.1cm]
{\footnotesize  \it Preprint: CP$^3$-Origins-2011-07}
 \end{abstract}

\maketitle


The Standard Model of high-energy physics provides a remarkably successful description of presently known phenomena. Yet, one sector of the Standard Model remains still to be confirmed. This is the one responsible for the generation of mass of all known elementary particles. The champion of this sector is the Higgs particle.  

It is well known that quantum corrections to the Higgs mass are proportional to the cutoff squared of the Standard Model. These corrections renders the model highly unnatural since a large fine tuning is needed to keep the mass of the Higgs at the electroweak scale. The large fine tuning is due to the fact that the cutoff scale can be as large as the Planck scale which is $10^{16}$ times larger than the electroweak scale. This is the famous hierarchy problem of the Standard Model. 

Two solutions stand out in the quest for a more natural Standard Model: Supersymmetry \cite{Dimopoulos:1981zb,Dimopoulos:1981yj} and Technicolor \cite{Weinberg:1979bn,Susskind:1978ms}. Supersymmetry renders scalars natural by linking them to their respective fermionic partners while in Technicolor the Higgs sector of the Standard Model is replaced by a new gauge dynamics featuring fermionic matter. Both solutions are still incomplete given that Supersymmetry requires the introduction of another sector needed to break it, while Technicolor needs to be extended to account for the masses of the Standard Model fermions.  

Here we put forward a novel solution to the hierarchy problem which makes use of gauge-gauge duality. We map a suitable minimal extension of the Standard Model into a new gauge theory featuring only fermionic matter. In other words the Standard Model is currently written in terms of {\it magnetic} fields and the dual {\it electric} description contains, besides gauge bosons, only fermionic degrees of freedom. We will elucidate our idea using  the investigations on gauge-gauge duality presented in \cite{Sannino:2009qc,Sannino:2009me,Sannino:2010fh,Mojaza:2011rw} and already used for relevant phenomenological predictions in \cite{Sannino:2010fh}.  This gauge-gauge duality framework was pioneered in a series of  ground breaking papers by Seiberg  \cite{Seiberg:1994bz,Seiberg:1994pq} for supersymmetric gauge theories. 

Quarks, leptons and the Higgs itself are degrees of freedom which are to be considered elementary in the magnetic description, but arise as composite states in the dual electric variables. The reason why it has been natural to introduce these states first when introducing the Standard Model is that the magnetic description is perturbative at the electroweak energy scale while the electric one is strongly coupled rendering harder the identification with the electrical variables. We will therefore make use of the fascinating possibility  that asymptotically free gauge theories have magnetic duals. 

Experimentally three generations of quarks and leptons have been discovered with the left handed fields transforming according to the doublet representation of the weak interactions and the right handed fields are singlets with respect to the weak interactions. Electromagnetic interactions are felt by left and right transforming electrically charged states. The Higgs sector of the Standard Model has not yet been experimentally established. To illustrate how our basic idea can work we start with a more symmetric looking Standard Model matter content, which might or might not be validated by future experiments,  and summarized in Table~\ref{SM}. 
\begin{table}[h]
\[ \begin{array}{|c| c | c c | } \hline
{\rm Fields} &  \left[ SU(3) \right] & SU(N_f)_L &SU(N_f)_R   \\ \hline \hline
q &\Yfund &{\overline{\Yfund }}&1 \\
\widetilde{q} & \overline{\Yfund}&1 &  {\Yfund}\\
l &1 &{\overline\Yfund }&1 \\
\widetilde{l} & 1&1 &  {\Yfund} \\
{H} & 1&\Yfund &  \overline{\Yfund}\\
 \hline \end{array} 
\]
\caption{We summarize here the Standard Model fermionic matter content. We have also generalized the Higgs field $H$. $SU(3)$ is the color gauge group and $N_f = 2n_g$ with $n_g$ the number of fermion generations.}
\label{SM}
\end{table}

We indicate with $q_{\alpha,{c}}^i$ the two component left spinor where $\alpha=1,2$
is the spin index, $c=1,...,3$ is the color index while
$i=1,...,N_f$ represents the flavor. $\widetilde{q}^{\alpha,c}_i$
is the two component conjugated right spinor. Similarly the leptonic fields are summarized in the table with $l_{\alpha}^i$ the two component left spinor and $\widetilde{l}^{\alpha,c}_i$ the two component conjugated right spinor. Here $N_f = 2 n_g$ with $n_g$ the number of Standard Model generations. The weak interactions are naturally embedded within the flavor group $SU(2n_g)_L \times SU(2n_g)_R$ by opportunely gauging $n_g$ times the $SU(2)_L \times U(1)_Y$ subgroup. The Higgs has been chosen to transform according to the bifundamental representation of $SU(2n_g)_L \times SU(2n_g)_R$ and therefore is not the minimal choice but it is the most natural one here. We set $SU(3)$ in between square brackets  in the table to indicate that this is the gauge group we are concentrating on to discuss how our gauge-gauge duality may work.  We therefore switch off the weak interactions for the time being.  {}From Table~\ref{SM} one is naturally led to consider the leptons as the forth color of an extended color group $ \left[ SU(4) \right] $. This is the renowned Pati-Salam  \cite{Pati:1973rp,Pati:1973uk,Pati:1974yy} extension of the Standard Model generalized to $N_f /2 = n_g$ generations. In Table~\ref{PS}  the spectrum of the Standard Model is summarized with respect to the Pati-Salam $SU(4)$  gauge group. 
\begin{table}[h]
\[ \begin{array}{|c| c | c c   | } \hline
{\rm Fields} &  \left[ SU(4) \right] & SU(N_f)_L &SU(N_f)_R    \\ \hline \hline
p &\Yfund &{\overline\Yfund }&1   \\
\widetilde{p} & \overline{\Yfund}&1 & {\Yfund}  \\
{H} & 1&\Yfund &  \overline{\Yfund}    \\
 \hline \end{array} 
\]
\caption{Fermion and Higgs matter content and their transformations with respect to the Pati-Salam SU(4) gauge group.}
\label{PS}
\end{table}
 We have now $p^i_{\alpha,C}$  ( $\widetilde{p}^i_{\alpha,C}$) with $C=1,2,3$ representing the ordinary left-handed quarks (conjugated right spinors) while $C=4$ are the leptons, and therefor $C$ is the vector index of the Pati-Salam group $SU(4)$. The $B-L$ symmetry is automatically embedded as one of the generators of $SU(4)$  \cite{Mohapatra:1980qe}.  This description of the Standard Model fields is, in practice, of book-keeping nature. To upgrade this model to a more realistic one Pati and Salam introduced several new  scalar degrees of freedom with the hope that one day there might be a more profound understanding of the origin behind the scalar sector.  Here, we will not duel with the specific details of the scalar potential and the pattern of chiral symmetry breaking. Our primary goal is to investigate if it is possible to construct a gauge dual of one of the simplest, and apparently unnatural, extensions of the Standard Model free from the hierarchy problem.  Therefore, we add only the minimum number of  matter fields allowing for such a possibility to manifest itself. We start by introducing the new complex scalars $\Phi_p$ ($\widetilde{\Phi}_{\widetilde{p}}$) transforming according to the fundamental (antifunamental) representation of the $ \left[ SU(4) \right] $ gauge group and fundamental (antifundamental) representation of the first (second) flavor group. 

At this point the spectrum of the theory is intriguingly close to the {\it magnetic} gauge dual envisioned in \cite{Mojaza:2011rw}. The states to add are a  magnetic Weyl fermion $\lambda_m$ transforming according to the adjoint representation of $\left[ SU(4) \right]$  and a Weyl fermion $M$ transforming as the Higgs with respect to the non-abelian flavor group. Adding these states leads to the spectrum reported in Table~\ref{PSE}. 
\begin{table}[h]
\[ \begin{array}{|c| c | c c c  c| } \hline
{\rm Fields} &  \left[ SU(4) \right] & SU(N_f)_L &SU(N_f)_R & U(1)_p & U(1)_{AF} \\ \hline \hline
\lambda_m &{\rm Adj} & 1 &1 &~~0 &~~1\\
p &\Yfund &{\overline\Yfund }&1&~~\frac{2n_g - 4}{4}&-\frac{4}{2n_g} \\
\widetilde{p} & {\overline\Yfund}&1 &  {\Yfund}& -\frac{2n_g - 4}{4} &-\frac{4}{2n_g}\\
\Phi_p &\Yfund &{\overline\Yfund }&1&~~\frac{2n_g - 4}{4} &-\frac{2n_g -4}{2n_g}\\
\widetilde{\Phi}_{\widetilde{p}} & {\overline\Yfund}&1 & {\Yfund}& -\frac{2n_g - 4}{4}&-\frac{2n_g -4}{2n_g}  \\
{M} & 1&\Yfund &  \overline{\Yfund}& ~~0&-1+\frac{8}{2n_g}   \\
{H} & 1&\Yfund &  \overline{\Yfund}& ~~0 &\frac{8}{2n_g}  \\
 \hline \end{array} 
\]
\caption{The high-energy complete magnetic spectrum including the fields of the Standard Model and their Pati-Salam extension.}
\label{PSE}
\end{table}
 We also make explicit the global symmetries of the new theory which are constituted by a new vector-like $U(1)_p$ and an axial one $U(1)_{AF}$ which is anomaly free. These global symmetries play a fundamental role via the 't Hooft anomaly conditions \cite{Hooft} in order to identify the correct electric theory.  By determining the most general set of solutions to these conditions, together with requiring consistent flavor decoupling and involution at the level of the electric and magnetic gauge groups in  \cite{Mojaza:2011rw}  we argued that the natural nonsupersymmetric electric dual theory is the one summarized in Table~\ref{dual}.
\begin{table}[t]
\[ \begin{array}{|c| c | c c c  c| } \hline
{\rm Fields} &  \left[ SU(2n_g - 4) \right] & SU(2n_g)_L &SU(2n_g)_R & U(1)_p & U(1)_{AF} \\ \hline \hline
\lambda &{\rm Adj} & 1 &1 &~~0 &~~1\\
P &\Yfund &{\Yfund }&1&~~1&-\frac{2n_g - 4}{2n_g} \\
\widetilde{P} & \overline{\Yfund}&1 &  \overline{\Yfund}& -1  &-\frac{2n_g - 4}{2n_g}\\
 \hline \end{array} 
\]
\caption{Electric dual of the magnetic Pati-Salam extension of the Standard Model whose spectrum is summarized in Table~\ref{PSE}.}
\label{dual}
\end{table}
In \cite{Mojaza:2011rw} we showed that it is possible to construct all the singlet states of the magnetic theory as composites of the electric ones. The only state we need to add to the table of \cite{Mojaza:2011rw} is $H$ which corresponds naturally to the electric composite gauge singlet $P\lambda \lambda \widetilde{P}$. It is remarkable that the electric dual theory does not contain scalar degrees of freedom. 

{A careful analysis of the phenomenological predictions of this model will be presented elsewhere. However, we can anticipate that since the magnetic extension of the Standard Model mimics the Pati-Salam one part of the phenomenological analysis will resemble to the one already present in the literature (see for example \cite{Toorop:2010yh}) }. 

Following \cite{Mojaza:2011rw} the dual {\it electric} gauge group is $SU(2n_g - 4) = SU(N_f  - 4)$. {}In order for the magnetic theory to be nonabelian we must have $2n_g - 4 \geq 2$ yielding the fundamental result that $n_g \geq 3$.  Of course, if $n_g=3$ the electric gauge group is $SU(2)$ and we expect an enhanced accidental global symmetry to occur i.e. $SU(2n_g)_L \times SU(2 n_g)_R \times U(1)_p  \subset SU(4 n_g)$, however if $n_g > 3 $ the electric theory has the same global symmetries of the magnetic one. Requiring the magnetic theory to remain asymptotically free we deduce the upper bound on the number of generations to be $6$ and therefore: 
\begin{equation}
3 \leq n_g \leq 6 \ .
\end{equation}
Interestingly duality can simultaneously render a Pati-Salam extension of the Standard Model natural and solve the mystery of the phenomenological existence of, at least, three generations of quarks and leptons. 

Our construction predicts the existence of few more matter fields around the energy scale where the Pati-Salam extended color gauge group $SU(4)$ appears. We expect this scale to be above  the TeV scale. The reader will recognize that our magnetic spectrum resembles a supersymmetric one, however,  the magnetic theory is not supersymmetric since we do not invoke supersymmetric relations among its spectrum and couplings  \cite{Mojaza:2011rw}. 

The dual electric theory is expected to be strongly coupled at the energy scale where the magnetic one is weakly coupled explaining why the quarks, the leptons and the Higgs seem elementary. Technically, the electric and magnetic descriptions are supposed to describe the same physics when both reach an infrared stable fixed point. Breaking of large distance conformality is therefore needed to ensure the occurrence of the phenomenologically viable pattern of spontaneous symmetry breaking. Re-instating the electroweak gauge interactions can, for example, de-stabilize the infrared fixed point but one could explore the introduction of new types  of  perturbations. In any event, the potential of the scalar sector should also emerge naturally in such a way to provide the correct patterns of breaking from $SU(4)$ to $SU(3)$ of color and the low energy decoupling of the magnetic degrees of freedom yet to be discovered. 

{ There are many ways to depart from conformality and some of them do not need the introduction of relevant operators. For example higher dimensional operators can achieve this. A well known example is the gauged Nambu Jona Lasinio model in which the theory can break conformality because of the introduction of four-fermion interactions (to be naturalized at a higher scale) \cite{Fukano:2010yv}.  Another possibility, as mentioned above, is the gauging of the weak interactions. The introduction of the new couplings are expected to modify the running of the couplings and eventually drive the theory away from the fixed points. Yet another possibility, as we have shown in \cite{Antipin:2011ny}, is to introduce renormalizable interactions modifying the structure of the fixed points.}

The Pati-Salam like extension of the Standard Model and our electric dual might not be unique and, in principle, an even more minimal magnetic extension of the Standard Model with associated electric description could exist.  

{ Some caveats  and further explanations are in order. Our results rely on the potential existence of gauge duals for nonsupersymmetric gauge theories featuring only fermionic matter whose dual involves both fermions and scalars. This type of duality has not yet been fully established for the theory presented here. There is, however, a known example in nature. This is ordinary QCD. The dual theory of QCD is the theory of hadrons which features composite fermions and scalars. Therefore nonsupersymmetric gauge theories can have dual gauge theories in a fashion similar to QCD of supersymmetric examples. Arguably the first test for the possible existence of gauge duals is the demonstration that the dual passes the 't Hooft anomaly matching conditions. {}For the theory used here in \cite{Mojaza:2011rw} it has been exhibit the infinite set of solutions of the 't Hooft anomaly conditions for any number of flavors and colors.  We then required a number of universal (i.e. susy independent) constraints. Among these criteria the most relevant ones are

 \begin{itemize}
 \item The involution condition. Dualizing the magnetic gauge group one recovers the electric one. The involution constraint has been proven in the appendix of \cite{Mojaza:2011rw}. There it was showed that the magnetic theory (susy ad nonsusy) must have a gauge group $SU(X)$ with $X = P(N_f) - N$ with N the number of colors of the electric theory. 
 
 \item Flavor decoupling. This condition requires and constrains the scalar spectrum for the magnetic dual theory. However this condition does not require the coupling constants of the dual theory to respect any supersymmetric condition. In fact gauge theories with identical UV spectrum but different couplings and potentials in the UV possess, typically, different infrared physics. 
 
 \item Consistency of the anomalous dimension constraints derived, for the electric theory, using only conformal invariance \cite{Sannino:2008nv,Sannino:2008pz}.

   \end{itemize}
The introduced scalars in the magnetic theory are composite states of the electric degrees of freedom in the same way that Seiberg's postulated dual quarks are composite states of the electric quarks. A famous example is QCD itself where many spin zero states emerge and are  instrumental to explain the low energy physics  of QCD.   

Furthermore in \cite{Antipin:2011ny} it has been shown that the magnetic dual theory envisioned in  \cite{Mojaza:2011rw}  possesses several infrared nonsupersymmetric fixed points. We have also shown the remarkable result that Seiberg's fixed point can emerge from nonsupersymmetric ultraviolet couplings couplings. These results allow us to be more confident about the envisioned duality since  they suggest that one can achieve infrared trustable fixed points with scalars and fermions without supersymmetry. Our result have been further confirmed by the recent analysis of Grinstein and Uttayarat in  \cite{Grinstein:2011dq}. These authors also claim that scalars without supersymmetry are light at fixed points. 

These recent results indirectly support the existence of gauge duals for nonsupersymmetric gauge theories supporting the envisioned realization of the Standard Model presented here.
}

Our results suggest the existence of a novel and elegant solution of the time-honored hierarchy problem in high-energy physics leading also to a new insight on the mystery of the experimental obervation of, at least, three generations of fundamental matter in nature.
\acknowledgments
We gladly thank Oleg Antipin, Simon Catterall, Phongpichit Channuie, Stefano Di Chiara, Tuomas Hapola, Jakob J. J\o rgensen, Chris Kouvaris, Matin Mojaza, Marco Nardecchia, Claudio Pica, Thomas Ryttov and Martin Svensson for interesting discussions, comments and careful reading of the manuscript.


\begin{thebibliography}{30}



\bibitem{Dimopoulos:1981zb}
  S.~Dimopoulos, H.~Georgi,
  Nucl.\ Phys.\  {\bf B193}, 150 (1981).

\bibitem{Dimopoulos:1981yj}
  S.~Dimopoulos, S.~Raby and F.~Wilczek,
  Phys.\ Rev.\  D {\bf 24}, 1681 (1981).

\bibitem{Weinberg:1979bn}
 S.~Weinberg,
 Phys.\ Rev.\  D {\bf 19}, 1277 (1979).

\bibitem{Susskind:1978ms}
 L.~Susskind,
 Phys.\ Rev.\  D {\bf 20}, 2619 (1979).


\bibitem{Sannino:2009qc}
  F.~Sannino,
  Phys.\ Rev.\  D {\bf 80}, 065011 (2009)
  [arXiv:0907.1364 [hep-th]].
 
\bibitem{Sannino:2009me}
  F.~Sannino,
  Nucl.\ Phys.\  B {\bf 830}, 179 (2010)
  [arXiv:0909.4584 [hep-th]].

\bibitem{Sannino:2010fh}
  F.~Sannino,
  Phys.\ Rev.\ Lett.\  {\bf 105}, 232002 (2010).
  [arXiv:1007.0254 [hep-ph]].

\bibitem{Mojaza:2011rw}
  M.~Mojaza, M.~Nardecchia, C.~Pica and F.~Sannino,
  arXiv:1101.1522 [hep-th]. Accepted for publication in Phys.~Rev.~D.

\bibitem{Seiberg:1994bz}
  N.~Seiberg,
  Phys.\ Rev.\  D {\bf 49}, 6857 (1994)
  [arXiv:hep-th/9402044].

\bibitem{Seiberg:1994pq}
  N.~Seiberg,
  Nucl.\ Phys.\  B {\bf 435}, 129 (1995)
  [arXiv:hep-th/9411149].

\bibitem{Pati:1973rp}
  J.~C.~Pati, A.~Salam,
  Phys.\ Rev.\ Lett.\  {\bf 31}, 661-664 (1973).
  
\bibitem{Pati:1973uk}
  J.~C.~Pati, A.~Salam,
  Phys.\ Rev.\  {\bf D8}, 1240-1251 (1973).

\bibitem{Pati:1974yy}
  J.~C.~Pati, A.~Salam,
  Phys.\ Rev.\  {\bf D10}, 275-289 (1974).
  
\bibitem{Mohapatra:1980qe}
  R.~N.~Mohapatra, R.~E.~Marshak,
  Phys.\ Rev.\ Lett.\  {\bf 44}, 1316-1319 (1980).

  \bibitem{Hooft} G. 't Hooft, {\it Recent Developments in Gauge Theories}, Plenum Press, 1980, 135; reprinted in {\it Unity of Forces in the Universe Vol. II}, A. Zee ed., World Scientific 1982, 1004.
 
 
\bibitem{Toorop:2010yh}
  R.~de Adelhart Toorop, F.~Bazzocchi, L.~Merlo,
  JHEP {\bf 1008}, 001 (2010).
  [arXiv:1003.4502 [hep-ph]]
 
 
\bibitem{Fukano:2010yv}
  H.~S.~Fukano, F.~Sannino,
  Phys.\ Rev.\  {\bf D82}, 035021 (2010).
  [arXiv:1005.3340 [hep-ph]].
 
\bibitem{Sannino:2008nv}
  F.~Sannino, R.~Zwicky,
  Phys.\ Rev.\  {\bf D79}, 015016 (2009).
  [arXiv:0810.2686 [hep-ph]].
  
\bibitem{Sannino:2008pz}
  F.~Sannino,
  Phys.\ Rev.\  {\bf D80}, 017901 (2009).
  [arXiv:0811.0616 [hep-ph]].
  
  
  
\bibitem{Antipin:2011ny}
  O.~Antipin, M.~Mojaza, C.~Pica, F.~Sannino,
  [arXiv:1105.1510 [hep-th]].
  
\bibitem{Grinstein:2011dq}
  B.~Grinstein, P.~Uttayarat,
  [arXiv:1105.2370 [hep-ph]].
  
\end{thebibliography}
\end{document}